\documentclass[aps,prd,showpacs,superscriptaddress]{revtex4}

\usepackage{amsmath,amsfonts,amssymb,graphicx,epsfig}

\allowdisplaybreaks[1]

\newcommand{\ie}{{\it i.e.}}

\newcommand{\eg}{{\it e.g.}}

\newcommand{\cf}{{\it cf.}}

\newcommand{\eq}{Eq.}

\newcommand{\fig}{Fig.}

\newcommand{\Ref}{Ref.}
\newcommand{\Refs}{Refs.}
\newcommand{\Sec}{Sec.}
\newcommand{\App}{App.}
\newcommand{\CP}{\emph{CP}}

\newcommand{\ket}[1]{\left|#1\right>}
\newcommand{\bra}[1]{\left<#1\right|}
\newcommand{\dd}[2]{\frac{{\rm d} #1}{{\rm d} #2}}
\renewcommand{\d}{{\rm d}}
\renewcommand{\i}{{\rm i}}
\DeclareMathOperator{\diag}{diag}
\DeclareMathOperator{\tr}{tr}

\begin{document}

\title{Neutrinos from WIMP Annihilations Obtained Using a Full Three-Flavor Monte Carlo Approach}

\author{Mattias Blennow}\email{emb@kth.se}
\affiliation{Department of Theoretical Physics,
School of Engineering Sciences, Royal Institute of Technology (KTH) --
AlbaNova University Center, Roslagstullsbacken 21, 106 91 Stockholm,
Sweden}

\author{Joakim Edsj{\"o}}\email{edsjo@physto.se}
\affiliation{Department of Physics, Stockholm University -- AlbaNova
University Center, Roslagstullsbacken 21, 106 91 Stockholm, Sweden}

\author{Tommy Ohlsson}\email{tommy@theophys.kth.se}
\affiliation{Department of Theoretical Physics,
School of Engineering Sciences, Royal Institute of Technology (KTH) --
AlbaNova University Center, Roslagstullsbacken 21, 106 91 Stockholm,
Sweden}

\begin{abstract}
Weakly Interacting Massive Particles (WIMPs) are one of the main
candidates for the dark matter in the Universe. If these particles
make up the dark matter, then they can be captured by the Sun or the
Earth, sink to the respective cores, annihilate, and produce
neutrinos. Thus, these neutrinos can be a striking dark matter
signature at neutrino telescopes looking towards the Sun and/or the
Earth. Here, we improve previous analyses on computing the neutrino
yields from WIMP annihilations in several respects. We include
neutrino oscillations in a full three-flavor framework as well as all
effects from neutrino interactions on the way through the Sun
(absorption, energy loss, and regeneration from tau decays). In
addition, we study the effects of non-zero values of the mixing angle
$\theta_{13}$ as well as the normal and inverted neutrino mass
hierarchies. Our study is performed in an event-based setting which
makes these results very useful both for theoretical analyses and for
building a neutrino telescope Monte Carlo code. All our results for the
neutrino yields, as well as our Monte Carlo code, are publicly
available. We find that the yield of muon-type neutrinos from WIMP
annihilations in the Sun is enhanced or suppressed, depending on the
dominant WIMP annihilation channel. This effect is due to an effective
flavor mixing caused by neutrino oscillations. For WIMP annihilations
inside the Earth, the distance from source to detector is too small to
allow for any significant amount of oscillations at the neutrino
energies relevant for neutrino telescopes.
\end{abstract}


\maketitle

\section{Introduction}

Weakly Interacting Massive Particles (WIMPs) are interesting (and
perhaps the most plausible) candidates for the dark matter in the
Universe. In particular, neutralinos (that arise naturally in supersymmetric extensions of the Standard Model) are promising WIMP candidates and
can be captured by \eg~the Sun or the Earth as the solar system moves
in the Milky Way halo \cite{cap-sun,cap-earth,lundberg-edsjo}. Once captured, they
will sink to the core and annihilate into several different
annihilation channels (leptons, quarks, gauge bosons, or Higgs
bosons). When the annihilation products decay, neutrinos of different
flavors are produced (in the case of \eg~Kaluza--Klein dark matter,
neutrinos can also be produced directly in the annihilation
processes \cite{KKnu}). During the propagation out of the Sun or the Earth, these
neutrinos can interact and undergo neutrino oscillations. Here, we
will improve on earlier estimates of the neutrino yields from WIMP
annihilations in the Sun and the Earth by performing a very detailed
analysis including both neutrino oscillations and interactions.

Neutrino oscillations will be included in a full three-flavor
framework including both vacuum and matter oscillations all the way
from the production region to a possible detector. Interactions are
only relevant for neutrinos propagating out of the Sun and we will
include absorption via charged-current (CC) interactions, energy
losses via neutral-current (NC) interactions, and regeneration of
neutrinos from tau decay (after CC interactions). Both neutrino
oscillations and interactions will be treated simultaneously in a
consistent way.

We will perform this analysis in an event-based framework, meaning
that we will follow each neutrino from production to detection. The
advantage of this approach is that, at the same time as we obtain
theoretical estimates of the neutrino yields at a neutrino telescope,
we will also produce events that can be used in Monte Carlo
simulations by the neutrino telescope community. In addition, our
code, which is written in the programming language Fortran, is
publicly available and can be implemented along with a neutrino
telescope Monte Carlo, which means that such a simulation is not
restricted to our finite list of parameter choices.

A recent analysis by Cirelli {\it et al.}~\cite{torino-osc} has been
performed along the same lines as our work. The main difference
between their work and our work is that they focus on the neutrino
flavor distributions based on a density matrix formalism and do not
follow each neutrino individually from the production point to the
detector. We will make comparisons with their results later in our
paper. In the final stages of preparation of this paper, we became
aware of the studies by Lehnert and Weiler \cite{Lehnert:2007fv} as
well as Barger {\it et al.}~\cite{Barger:2007xf}. In the former, the
neutrino flavor flux ratios from WIMP annihilations in the Sun in the
absence of neutrino interactions are presented (which is valid up to
neutrino energies of the order of 10~GeV, above which neutrino interactions affect the fluxes by more than a few percent). We will find that our low-energy results agree very well with those of Lehnert and Weiler, while our high-energy results differ due to our inclusion of neutrino interaction effects. In the latter, the
approach is similar to that of Cirelli {\it et al.}, but also includes
the dependence on the spin of the WIMPs and mainly discusses annihilation channels where equal amounts of all neutrino flavors are produced, and thus, the neutrino oscillation effects are suppressed. Furthermore, our study also includes neutrinos created by WIMP annihilations in the Earth. In addition, Crotty \cite{Crotty:2002mv} has earlier published the results of using a similar Monte Carlo code when treating the annihilation of supermassive ($10^8$~GeV $< m_\chi < 10^{16}$~GeV), strongly interacting dark matter particles.

This paper is organized as follows. In \Sec~\ref{sec:capann}, we will
review how WIMPs are captured in the Sun or the Earth and how
annihilations are treated. Next, in \Sec~\ref{sec:interosc}, we will
describe our framework for including neutrino interactions and
oscillations into our Monte Carlo code. Then, in sections
\ref{sec:prop}-\ref{sec:propearth}, we will discuss how we perform the
neutrino propagation out of the Sun, from the Sun to the Earth, and
through the Earth to an actual detector. The intermediate results for
the neutrino yields of different flavors at the surface of the Sun, at
a distance of 1~AU, and at the actual detector will be presented. In
\Sec~\ref{sec:nonzero_comparison}, we will investigate the impact of
non-zero $\theta_{13}$ as well as that of having normal or inverted
neutrino mass hierarchy, and in \Sec~\ref{sec:anniEarth}, we will
consider WIMP annihilations in the Earth. Finally, in
\Sec~\ref{sec:conclusions}, we will summarize our results and present
our conclusions. In addition, in the appendix, we will show the
equivalence of our Monte Carlo simulation and the density matrix
formalism (used in \eg~\Ref~\cite{torino-osc}).

\section{Annihilation of WIMPs in the Sun/Earth}
\label{sec:capann}

WIMPs can be captured in the Sun \cite{cap-sun} or the Earth
\cite{cap-earth} when they move through the dark matter halo in the
Milky Way. Once the WIMPs have been captured, they will sink to the
core of the Sun or the Earth, where they can annihilate and produce
leptons, quarks, gauge and Higgs bosons. These will hadronize, decay,
and eventually, produce neutrinos. We therefore follow the
calculations performed in \Refs~\cite{jenutel,jenutel2}, but update
them to be more general and include more annihilation channels. We use
{\sc Pythia 6.400} \cite{pythia} to simulate the hadronization and
decay of the annihilation products and collect the neutrinos and
antineutrinos produced. The annihilation channels that we simulate are
the following: $c \bar{c}$, $b
\bar{b}$, $t \bar{t}$, $\tau^+ \tau^-$, $W^+ W^-$, $Z^0 Z^0$, $g g$,
$\nu_e \bar{\nu}_e$, $\nu_\mu \bar{\nu}_\mu$, and $\nu_\tau
\bar{\nu}_\tau$ \footnote{Lighter quarks can be simulated with our code,
but do not give any significant neutrino yields, and hence, we do not include them in our simulations.}, where the last three channels are only relevant for
WIMP candidates that can annihilate directly to neutrinos
(\eg~Kaluza--Klein dark matter). The lighter charged leptons are not
of importance, since electrons are stable and muons are stopped before
they have a chance to decay and produce (high-energy) neutrinos. For
the $b \bar{b}$ channel, we need to be especially careful, since these
hadronize and produce $B$ mesons, which, for the case of the Sun,
interact before they decay. We include these interactions in an
approximate fashion by performing the {\sc Pythia} simulations as if
in free space, and later, we rescale the energy of the resulting
neutrinos by estimating the energy loss of the $B$ mesons due to their
interactions. This follows the calculation in \Ref~\cite{RS}, but we
use newer estimates of the $B$ meson interaction cross-sections (as given in \Ref~\cite{jephd}). For
the solar density, we use the Standard Solar Model
\cite{bs05}. Annihilation channels including Higgs bosons are not
simulated directly, since we can easily calculate the neutrino yield
from the decay of the Higgs bosons given the Higgs masses and decay
branching rates. For a specific WIMP model, we later perform this step
separately, once the Higgs properties are known. In this calculation,
we let the Higgs bosons decay in flight, properly integrating over the
decay angles and performing the appropriate Lorentz boost to obtain
the neutrino yields in the rest frame of the Sun or the Earth. We
perform simulations for the following WIMP masses: 10, 25, 50, 80.3,
91.2, 100, 150, 176, 200, 250, 350, 500, 750, 1000, 1500, 2000, 3000,
5000, and 10000~GeV. For each mass and annihilation channel, we
perform $2.5 \times 10^6$ annihilations and collect all neutrinos and
antineutrinos, keeping track of the flavors separately. In the next
section, we will describe our formalism for neutrino interactions and
oscillations, and in the succeeding sections, we will discuss how we
propagate the neutrinos from the core of the Sun and/or Earth to the
detector.

\begin{figure}
\begin{center}
\epsfig{file=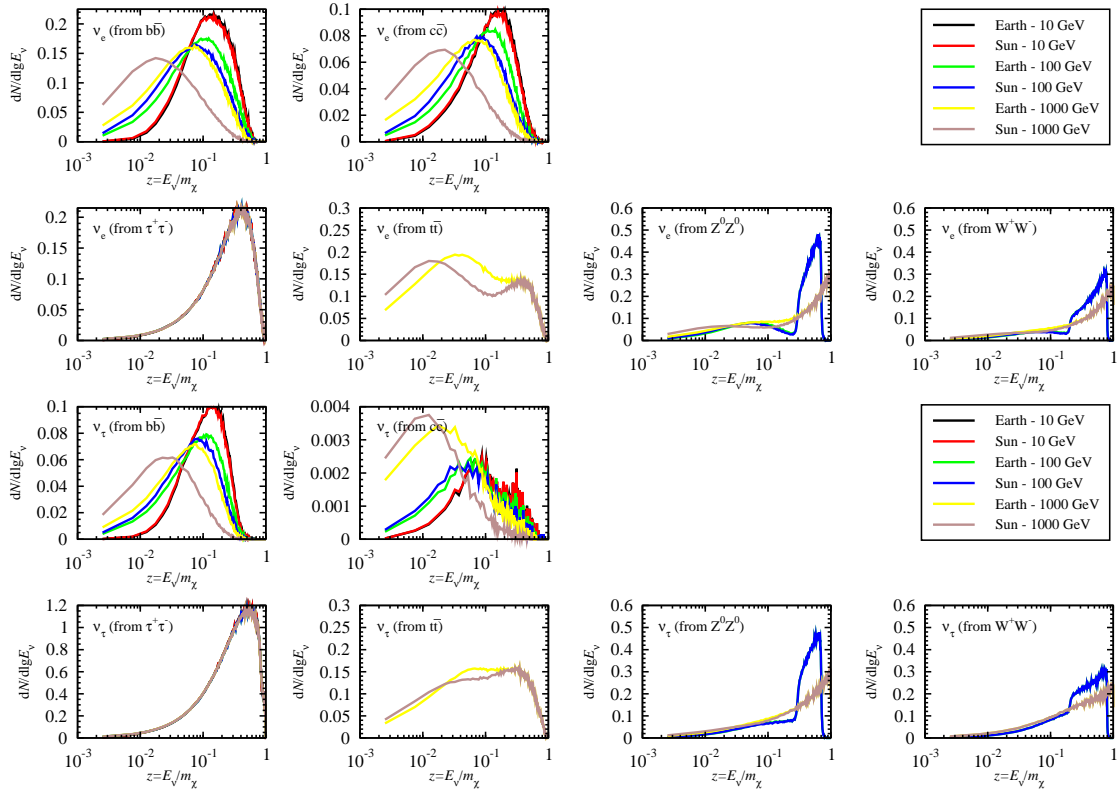,angle=270,width=\textwidth}
\vspace{-1.5cm}
\caption{The neutrino yields for electron and tau neutrinos as
functions of $z = E_\nu/m_\chi$ for six different WIMP annihilation
channels at production in the center of the Sun and the Earth. Note that the muon neutrino yields are the same as the electron neutrino yields and are therefore not shown separately.}
\label{fig:production}
\end{center}
\end{figure}

In \fig~\ref{fig:production}, we present our results for the neutrino
yields at creation in some of the different annihilation channels.
Comparing this to \fig~2 in the published version of
\Ref~\cite{torino-osc}, we find that our results for the $W^+ W^-$
annihilation channel is about a factor of two lower and that there is
a discrepancy in the $t\bar t$ channel, where we also have a lower
yield. We also see a double-peaked yield for the $t \bar{t}$ channel
compared to the broader single-peaked yield in
\Ref~\cite{torino-osc}. In addition, our high-energy peak goes up to
higher energies than their results. This two-peak structure is
expected as the high-energy peak comes from the prompt decay of $W$
bosons and the low-energy peak comes from quark jets.

We have been in contact with the authors of \Ref~\cite{torino-osc}
regarding these differences, and as a consequence of this, they have
found two errors in their code. The first one was a factor of two too
high yields for the $W^+W^-$ channel (which affected the yield from
$W^+W^-$ channel directly and the yield from $t\bar{t}$
indirectly). The other one was an error in the Lorentz boost of the
$W$ boson resulting from top decay. After the authors of
\Ref~\cite{torino-osc} corrected these errors, our yields now agree
reasonably well \footnote{The authors of \Ref~\cite{torino-osc} will publish an erratum for their paper and update the data available on their website regarding their work.}. There is still a small difference in the $t \bar{t}$
channel, but this difference is mostly due to our inclusion
of final state radiation of gluons.

\section{Neutrino interactions and oscillations}
\label{sec:interosc}

On their way from creation in the center of the Sun or the Earth to
measurement in the detector, neutrinos can undergo both
neutrino-nucleon interactions and neutrino oscillations. These are
treated according to the following two prescriptions.

\subsection{Neutrino interactions}

Neutrinos can undergo both CC and NC interactions with nucleons during
their passage through a medium. In a CC interaction, we obtain a
charged lepton, whereas in a NC interaction, we obtain a neutrino with
degraded energy. We calculate the total and differential
cross-sections for these interactions, as both are needed to determine
the absolute interaction rates and the energy loss of the lepton. We
need the latter to take into account energy losses of neutrinos in the
Sun and regeneration of neutrinos from tau decay, as well as for
calculating the scattering process near or in the neutrino detector.
We use the convenient expressions for the (anti)neutrino-nucleon
cross-sections that can be found in \Ref~\cite{levy}. These
expressions include the effect of the tau mass on the
(anti)neutrino-nucleon CC cross-sections.

In order to calculate the cross-sections, we need to specify the
parton distribution functions (PDFs). We use the CTEQ6 \cite{cteq6}
PDFs, or more specifically, the CTEQ6-DIS PDFs. For the other
physical parameters, we choose the following values
\begin{eqnarray*}
  M_W & = & 80.41 \mbox{~GeV}, \\
  M_Z & = & 91.188 \mbox{~GeV}, \\
  G_F & = & 1.16639 \times 10^{-5} \mbox{~GeV$^{-2}$}, \\
  M_p & = & 0.938272 \mbox{~GeV}, \\
  M_n & = & 0.939566 \mbox{~GeV}, \\
  \sin^2 \theta_W & = & 0.23124.
\end{eqnarray*}
The resulting cross-sections agree well with other calculations in
this energy range, see \eg~\Ref~\cite{gqrs98} (however, only
cross-sections on isoscalar targets are given in that work).

The neutrino-nucleon CC cross-sections have been measured by several
experiments (see \Ref~\cite{pdg} for a summary). In the energy range
up to $E_\nu = 350$~GeV, the average cross-sections are \cite{pdg}
\begin{eqnarray*}
  \frac{\sigma_{\nu N}}{E_\nu} & = & (0.677 \pm 0.014) \times 10^{-38}
  \mbox{~cm$^2$/GeV}, \\
  \frac{\sigma_{\bar{\nu} N}}{E_\nu} & = & (0.334 \pm 0.008) \times
  10^{-38} \mbox{~cm$^2$/GeV}.
\end{eqnarray*}
As a comparison, our calculated cross-sections at 100~GeV are
\begin{eqnarray*}
  \frac{\sigma_{\nu N}}{E_\nu= 100 \mbox{~GeV}} & = & 0.684 \times
  10^{-38} \mbox{~cm$^2$/GeV}, \\
  \frac{\sigma_{\bar{\nu} N}}{E_\nu= 100 \mbox{~GeV}} & = & 0.330
  \times 10^{-38} \mbox{~cm$^2$/GeV},
\end{eqnarray*}
\ie, very close to the measured cross-sections. 

Note that, in order to gain computational speed, we do not use the
full expressions for the total cross-sections directly, instead we
calculate the different cross-sections for a range of energies from
1~GeV to $10^{12}$~GeV and interpolate these results. The
interpolation errors introduced are below 1~\%, but the speed increase
is dramatic.

In order to take care of each neutrino-nucleon interaction, we have
developed a Monte Carlo code that simulates the interaction and
returns the energy and angles of the final state lepton as well as the
hadronic shower. The neutrino-interaction and Monte Carlo codes are
available for download at the website in \Ref~\cite{wimpsim}, where
more technical information is also available.

Our differential spectra differ somewhat from those obtained with \eg~{\sc Pythia}. The main reason is that {\sc Pythia} does not
allow us to use the full phase-space (\ie, going down to sufficiently
low momentum transfer $Q^2$). This is not surprising, since this is
not a process for which {\sc Pythia} is optimized. We can also observe
this effect looking at the total cross-sections, which {\sc Pythia}
underestimates at low energies (less than 100~GeV), again due to the
fact that the phase-space is not complete.

\subsection{Neutrino oscillations}
\label{sec:nuosc}

We want to describe the yields of neutrinos at the Earth as accurately
as possible. In order to accomplish this task, we need to include the
effects of neutrino oscillations, which have been observed in
experiments
\cite{Fukuda:1998mi,Fukuda:2002pe,Ahmad:2002jz,Ahmad:2001an,Eguchi:2002dm,Araki:2004mb,Ahn:2002up,Aliu:2004sq,Michael:2006rx,Ashie:2004mr}. For
the high-energy neutrinos that we are considering, the adiabatic
propagation of neutrino matter eigenstates inside the Sun, which is
used for ordinary solar neutrinos, is no longer valid. The reason is
mainly that the neutrinos are produced at densities which are larger
than the high-energy resonance density, corresponding to the resonance
of the small vacuum mixing angle $\theta_{13}$. In addition, for
certain setups (most notably for high-energy neutrinos), it may not be
a good idea to use the approximation of decoherent mass eigenstates
arriving at the Earth, which is why we store these neutrinos on an
event basis rather than as energy spectra for the different mass
eigenstates.

The theory of neutrino oscillations is based on the assumption that
the neutrino flavor eigenstates are not equivalent to the neutrino
mass eigenstates, but rather that they are linear combinations of each
other. This can be written as
\begin{equation}
\ket{\nu_\alpha} = \sum_a U_{\alpha a}^* \ket{\nu_a},
\end{equation}
where $U$ is the leptonic mixing matrix, $\ket{\nu_\alpha}$ is a
neutrino flavor eigenstate, and $\ket{\nu_a}$ ($a = 1,2,3$) are the
neutrino mass eigenstates with definite masses $m_a$. The evolution
of the neutrino state $\nu(t)= (\nu_e(t), \nu_\mu(t), \nu_\tau(t))^T$
is given by
\begin{equation}
\nu(t) = S(t) \nu(0),
\end{equation}
where the evolution operator $S(t)$ depends on the distance and medium
traversed. For three-flavor neutrinos propagating in matter of
constant density, the evolution operator is given by $S(t) =
\exp(-{\rm i}H t)$, where $H$ is the total Hamiltonian including both
the vacuum and the Mikheyev--Smirnov--Wolfenstein (MSW) potential
terms, \ie,
\begin{equation}
H = \frac{1}{2E} U\diag(0,\Delta m_{21}^2,\Delta m_{31}^2) U^\dagger
+ \diag(\sqrt{2}G_F N_e,0,0).
\end{equation}
Here $E$ is the neutrino energy, $\Delta m_{ij}^2 = m_i^2-m_j^2$ are
the neutrino mass squared differences, $G_F$ is the Fermi coupling
constant, and $N_e$ is the electron number density. By the use of the
Cayley--Hamilton formalism, it is possible to write the evolution
matrix as \cite{Ohlsson:1999xb}
\begin{equation}
S(t) = \phi \sum_{a=1}^3 
e^{-{\rm i}\lambda_a t} 
\frac{(\lambda_a^2+c_1)I+\lambda_a T+T^2}{3\lambda_a^2+c_1},
\label{eq:sch}
\end{equation}
where $\phi$ is an overall phase factor which does not affect neutrino
oscillations, $I$ is the $3 \times 3$ identity matrix, $T = H - \tr(H)
I/3$ is the traceless part of the Hamiltonian, $c_1 = -\tr(T^2)/2$,
and $\lambda_a$ are the eigenvalues of $T$ [which can be determined
from $c_1$ and $\det(T)$]. The probability of an initial neutrino of
flavor $\nu_\alpha$ to be in the flavor eigenstate $\nu_\beta$ at time
$t$ is then given by $P_{\alpha\beta}(t) = |S_{\beta\alpha}(t)|^2$.

When propagating neutrinos through matter of varying density, we can
approximate the electron number density profile by a large number of
layers with constant electron number density (the appropriate number
of layers is determined by the rate of change in the electron number
density as well as the neutrino oscillation lengths). If we let a
neutrino pass through $k$ layers and label the evolution operator of
layer $i$ by $S_i$, then the total evolution operator $S$ is given by
\cite{Ohlsson:1999um}
\begin{equation}
S = S_k S_{k-1} \ldots S_2 S_1.
\label{eq:seq}
\end{equation}
Neutrino oscillations in vacuum can also be treated with this method,
we simply use one layer with the appropriate length and set the
electron number density to zero. We will later calculate $S$ with
these methods using accurate matter density profiles for both the Sun
and the Earth. Note that the calculation of $S$ in \eq~(\ref{eq:seq}) is particularly
simple given the Cayley--Hamilton formalism presented in \eq~(\ref{eq:sch}).

The neutrino oscillation probabilities only depend on the leptonic
mixing matrix and the mass squared differences. The standard
parameterization of the leptonic mixing matrix is given by
\cite{Yao:2006px}
\begin{equation}
U = \left(\begin{array}{ccc}
c_{13}c_{12} & c_{13}s_{12} & s_{13}e^{-{\rm i}\delta} \\
-s_{12}c_{23} - c_{12}s_{23}s_{13}e^{{\rm i}\delta} &
c_{12}c_{23} - s_{12}s_{23}s_{13}e^{{\rm i}\delta} & s_{23}c_{13} \\
s_{12}s_{23} - c_{12}c_{23}s_{13}e^{{\rm i}\delta} &
-c_{12}s_{23} - s_{12}c_{23}s_{13}e^{{\rm i}\delta} & c_{23}c_{13}
\end{array}\right),
\end{equation}
where $c_{ij} = \cos(\theta_{ij})$, $s_{ij} = \sin(\theta_{ij})$, and $\delta$ is the $CP$-violating phase. The
present phenomenological constraints on the neutrino oscillation
parameters are given by global fits to the different neutrino
oscillation experiments as \cite{Maltoni:2004ei}
\begin{align}
& \theta_{12} = 33.2^\circ \pm 4.9^\circ, \nonumber\\
& \theta_{13} < 12.5^\circ, \nonumber\\
& \theta_{23} = 45.0^\circ \pm 10.6^\circ, \nonumber\\
& \delta \in [0,2\pi), \nonumber\\
& \Delta m_{21}^2 = (8.1^{+1.0}_{-0.9}) \times 10^{-5} \, {\rm eV}^2,
    \nonumber\\
& |\Delta m_{31}^2| = (2.2^{+1.1}_{-0.8}) \times 10^{-3} \, {\rm eV}^2,
    \nonumber
\end{align}
where the presented values are the best-fits and the $3\sigma$
(99.7~\% confidence level) ranges. The leptonic mixing angle
$\theta_{12}$ and the small mass squared difference $\Delta m_{21}^2$
are mainly constrained by solar and long-baseline reactor experiments,
while the leptonic mixing angle $\theta_{23}$ and the large mass
squared difference $\Delta m_{31}^2$ are mainly constrained by
experiments studying atmospheric and accelerator neutrinos. The upper
bound on the leptonic mixing angle $\theta_{13}$ is currently given by
short-baseline reactor experiments, while the \CP-violating phase $\delta$ is
completely unknown.

In our simulations, we have focused on the best-fit values for
$\theta_{12}$, $\theta_{23}$, $\Delta m_{21}^2$, $|\Delta m_{31}^2|$,
and put the \CP-violating phase $\delta$ equal to zero. The different
combinations of the following scenarios were implemented:
\begin{itemize}
\item Three different values of $\theta_{13}$ ($\theta_{13} = 0,
5^\circ, 10^\circ$ corresponding to $\sin^2 2\theta_{13} = 0, 0.03,
0.12$, respectively)
\item Normal or inverted mass hierarchy (\ie, $\Delta m_{31}^2$
positive or negative)
\end{itemize}
In addition, we have implemented the case when there are no neutrino
oscillations by setting $\theta_{ij} = 0$ and $\Delta m_{ij}^2 =
0$. The reason is to study the effect of neutrino oscillations
separately as well as being able to compare our non-oscillation
results with previous works that have not taken neutrino oscillations
into account. For the remainder of this paper, the examples shown are
for $\theta_{13} = 0$ and normal neutrino mass hierarchy unless stated
otherwise (\cf, \Sec~\ref{sec:nonzero_comparison}). Thus, our standard
set of neutrino oscillation parameters (std.~osc.) is $\theta_{12} =
33.2^\circ$, $\theta_{13} = 0$, $\theta_{23} = 45^\circ$, $\delta =
0$, $\Delta m_{21}^2 = 8.1 \times 10^{-5}$~eV$^2$, and $\Delta
m_{31}^2 = 2.2 \times 10^{-3}$~eV$^2$.

\section{Propagation out of the Sun}
\label{sec:prop}

We are now ready to perform the steps necessary to propagate the
neutrinos produced at the center of the Sun or the Earth calculated in
\Sec~\ref{sec:capann}. We start with the Sun, as it is more involved,
and take care of the Earth later.

For the Sun, we divide the propagation to the detector into several
steps, starting with the propagation from the center of the Sun to its
surface. In this process, we take into account both neutrino
interactions (CC and NC) as well as neutrino oscillations.

\subsection{Solar model}

Let us first describe the solar model that we have used. We need a
good knowledge of the Sun's interior in order to know both the matter
effects on neutrino oscillations and the interaction probabilities. We
have used the Standard Solar Model \cite{bs05}, which gives the
density of the dominant elements and electrons as a function of the
radial distance from the center of the Sun. We use the element
densities to calculate the density of protons and neutrons as a
function of radius, which is needed for the neutrino interaction
probabilities, and use the electron densities to calculate the matter
potential needed for the neutrino oscillation matter effects.

\subsection{Neutrino interactions and oscillations}

Since neutrino interactions and oscillations can occur simultaneously,
we need to take both into account at the same time. However, the
neutrino-nucleon cross-sections are practically flavor independent,
which means that this is particularly easy to perform. (This is true
as long as the tau mass can be neglected. We include the effects of
the tau mass later on when we treat the tau neutrino CC interactions.)
For each neutrino, we first calculate a creation point. Since the
WIMPs are in thermal contact with the Sun's core, they will be
distributed according to a Gaussian distribution
\begin{equation} \label{eq:aprof1}
  n(r) = n(0) {\rm e}^{-r^2/(2r_{\chi}^2)}
\end{equation}
with
\begin{equation} \label{eq:aprof2}
  r_{\chi} = \left( \frac{3kT}{4 \pi G \rho m_\chi} \right)^{1/2},
\end{equation}
where $k$ is the Boltzmann constant, $T$ is the core temperature in
the Sun, $G$ is the gravitational constant, $\rho$ is the core
density, and $m_\chi$ is the WIMP mass. For the Sun, annihilations are
concentrated to within about 1~\% of the solar radius, which means
that the effect is small. We choose a creation point and calculate the
path length from this creation point to the surface of the Sun. Due to
the concentration of the annihilations to the very core, we can safely
approximate this path as being purely radial to simplify our calculations.
After this stage, we
know how much material the neutrino has to traverse in order to reach
the solar surface. We then compute the neutrino-nucleon cross-section
and use it to randomize the point of interaction for each neutrino (or
if the neutrino escapes the Sun without interacting). Depending on the
type of interaction (CC or NC), we simulate the interaction results.

In the case of a CC interaction, we use the neutrino evolution
operator method described in \Sec~\ref{sec:nuosc} to determine the
probabilities of the neutrino being in the different flavor
eigenstates at the time of interaction. The actual interacting
neutrino flavor is then randomized according to these probabilities.
If the neutrino at this interaction point is a tau (anti)neutrino, we
simulate the neutrino interaction (properly including the tau mass
suppression of the tau (anti)neutrino-nucleon cross-section \footnote{What we actually do
is to simulate the CC interactions with equal cross-sections for all flavors (since we do not know the flavor before-hand). Next, we check if it is a tau neutrino in which case we renormalize the cross-section. Then, if it happens that the interaction should not have taken place, we put the neutrino back at that position in the Sun without projecting out the state and continue with the propagation.}). We then
obtain a charged tau lepton, whose decay is simulated with {\sc
Pythia}. We collect the neutrinos produced in this decay and proceed
with their propagation out of the Sun. If the neutrino is an electron
or muon neutrino, then the charged lepton produced in the CC
interaction is either stable or will be stopped before it has had time
to decay and the neutrino is considered to be absorbed.

The main effect of a NC interaction is to change the energy of the
neutrino. Since the NC vertex is flavor-blind, it does not destroy the
relative phases among different neutrino flavor eigenstates and does
not force the neutrino to interact as a particular flavor (see
\eg~\Ref~\cite{Smirnov:1991eg} for a discussion on the oscillations of
neutrinos from $Z^0$-decay, the argument can be readily extended to NC
interactions of neutrinos). After an interaction, we then repeat this
procedure until the neutrino has reached the surface of the Sun. In
\App~\ref{app:MCrho-equiv}, it is shown that the Monte Carlo method
described above is statistically equivalent to the density matrix
formalism used in \Ref~\cite{torino-osc}.

For the neutrino oscillations between creation and interaction points,
we use the method described in \Sec~\ref{sec:nuosc}. This means that
we divide the Sun, with its varying matter density, into layers of
constant density. We have chosen to let each layer have a width of
0.3~\% of the radius of the Sun. This gives a total error on the
neutrino state after propagation of much less than 1~\%, which is a
reasonable compromise between accuracy and speed.

\subsection{Results for neutrino yields at the surface of the Sun}

\begin{figure}
\centerline{\epsfig{file=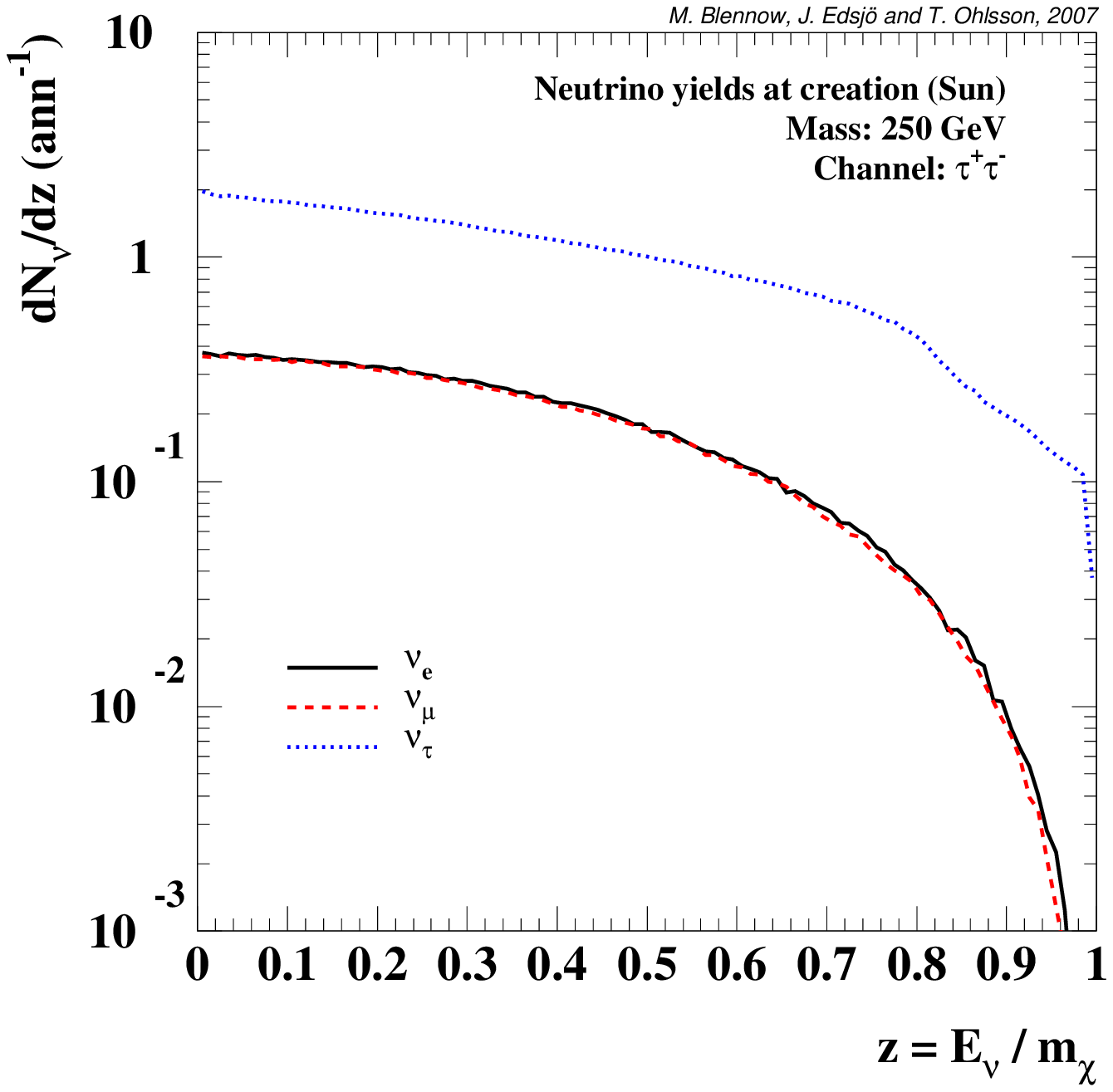,width=0.49\textwidth}
\epsfig{file=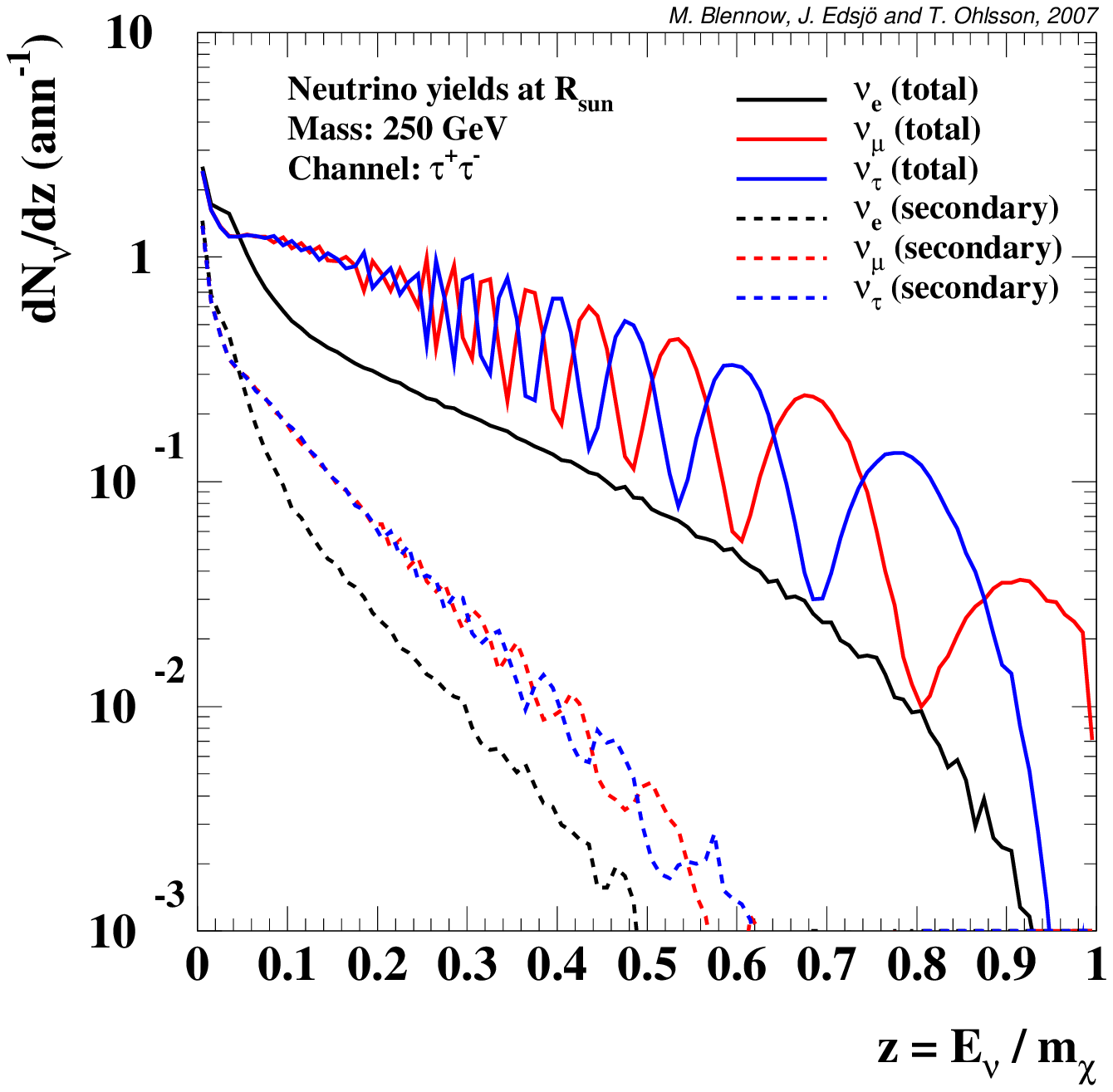,width=0.49\textwidth}}
\caption{The neutrino yields as a function of $z = E_\nu/m_\chi$ at
creation in the center of the Sun (left panel) and at the surface of
the Sun (right panel) for annihilation of 250~GeV WIMPs into $\tau^+
\tau^-$. The secondaries in the right-hand panel are neutrinos that come from $\tau^\pm$ decay after a $\nu_\tau /\bar{\nu}_\tau$ CC interaction.}
\label{fig:tau-rsun}
\end{figure}

We are now ready to present some results at the solar surface in order
to show the effects of propagation through the Sun. Since most
annihilation channels produce similar amounts of neutrinos of
different flavors, we focus on annihilation to $\tau^+ \tau^-$, which
produces far more tau neutrinos than electron and muon neutrinos, in
order to emphasize the effects of neutrino oscillations. In
\fig~\ref{fig:tau-rsun}, we show the yields of neutrinos from
annihilation of 250~GeV WIMPs to $\tau^+ \tau^-$ both at the center
and at the surface of the Sun. This figure is derived for the standard
set of neutrino oscillation parameters as described earlier.
\begin{figure}
\centerline{\epsfig{file=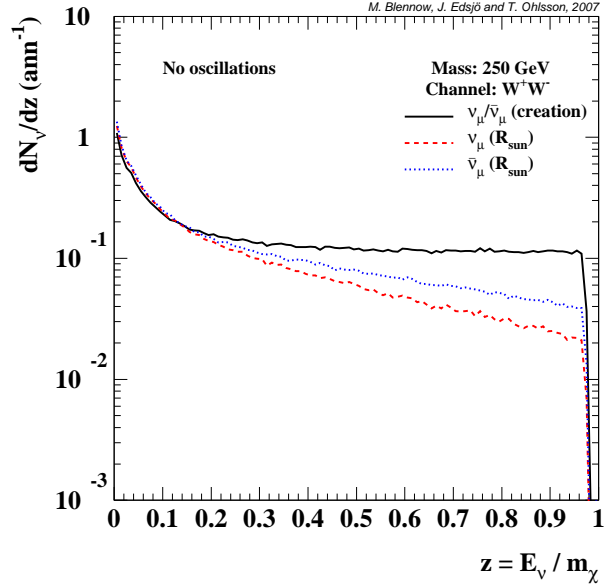,width=0.49\textwidth}}
\caption{The muon neutrino and antineutrino yields as a function of $z
= E_\nu/m_\chi$ at creation and at the surface of the Sun for
annihilation of 250~GeV WIMPs into $W^+ W^-$. In this figure, neutrino
oscillations have been turned off in order to show the effect of
neutrino interactions.}
\label{fig:w-no-osc}
\end{figure}
We can clearly observe two main effects in this figure.

The first effect is that oscillations effectively mix muon and tau
neutrinos on their way out of the Sun, whereas the electron neutrinos
remain essentially unmixed. This is due to the electron neutrino being
equivalent to one of the matter eigenstates when propagating in a
medium of high electron number density (which is true for the most
part of the neutrino propagation inside the Sun). The almost complete
mixing of muon and tau neutrinos is due to the effective mixing angle
in the remaining two-flavor system being very close to $\theta_{23}$
\cite{Blennow:2004js}, which is maximal in our simulations, while the
dependence on $\theta_{12}$ is suppressed by the ratio $\alpha =
\Delta m_{21}^2/\Delta m_{31}^2$.

The second effect is the pile-up of events at low energies. Some of
these events come from secondary decays of $\tau^\pm$ that have been created in
$\nu_\tau /\bar{\nu}_\tau$ CC interactions and some of these come
from energy losses due to NC
interactions. The interactions also create a loss at higher energies
(both due to absorption via CC interactions of electron and muon
neutrinos and due to energy losses). This is not fully visible in this
figure, but in \fig~\ref{fig:w-no-osc}, we can observe this effect
more clearly for the $W^+W^-$ channel with neutrino oscillations
turned off. The loss of neutrinos at higher energies due to these
interactions is clearly visible.

\section{Propagation from the Sun to the Earth}
\label{sec:vacprop}

When the neutrinos have escaped from the Sun, we use the neutrino
evolution operator method described in \Sec~\ref{sec:nuosc} to
propagate the neutrinos from the surface of the Sun to a distance of
1~AU from the Sun. In the treatment of the oscillations of ordinary
solar neutrinos (produced by the thermonuclear reactions sustaining
the Sun), it is common practice to assume that the neutrinos arrive at
the Earth as mass eigenstates. The main arguments for making this
assumption are the following:
\begin{itemize}
\item The coherence length due to separation of wave packets for the
neutrino mass eigenstates is much shorter than 1~AU.
\item The change in the neutrino baseline due to the diameter of the
Earth and the eccentricity of the Earth's orbit is comparable to or
larger than the oscillation length and also gives an effective average
of the oscillations.
\item The finite energy resolution of detectors results in an
effective averaging of the resulting fast oscillations.
\end{itemize}
The first of these assumptions is no longer valid when we study the
high-energy neutrinos produced by WIMP annihilations. Thus, we save
the neutrinos arriving at 1~AU with both energy and relative phase
information among the neutrino flavor eigenstates (in order to make it
possible to propagate the neutrinos further), rather than making
energy spectra for the different flavors. Depending on the actual
neutrino flux, neutrino telescopes may be sensitive to its temporal
variation, which essentially would render the second assumption
invalid (we will come back to this issue later).
However, the energy resolution of a typical neutrino
telescope is still not very high, making it plausible that the third
assumption could still hold (in which case we could just compute the
energy spectra for the different neutrino mass eigenstates arriving at
the Earth). However, in some dark matter scenarios, such as
Kaluza--Klein dark matter, the WIMPs may annihilate directly into
pairs of neutrinos and antineutrinos, resulting in a sharp
monochromatic peak in the spectrum. The smudging of the energy
spectrum from the poor energy resolution of the detector does not
change the neutrino oscillation probabilities of such monochromatic
neutrinos, and thus, we keep the most general setup in order for our
method to be independent of the type of dark matter that is studied.
The results after this stage are stored on an event-by-event basis
complete with neutrino energies, amplitudes and phases.

\subsection{Results for neutrino yields at 1~AU}

\begin{figure}
\centerline{\epsfig{file=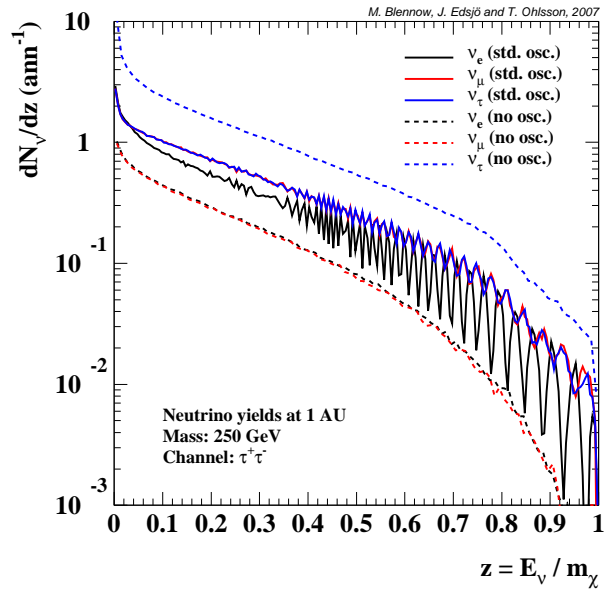,width=0.49\textwidth}}
\caption{The neutrino yields as a function of $z = E_\nu/m_\chi$ at
1~AU for annihilation in the Sun of 250~GeV WIMPs into $\tau^+ \tau^-$
both with (std.~osc.) and without neutrino oscillations. The main
effects of oscillations are easily seen, \ie, muon and tau neutrinos
are fully mixed, and electron neutrinos are partially mixed with the
other flavors.}
\label{fig:tau-1au}
\end{figure}

The vacuum neutrino oscillations from the Sun to the detector force
the electron neutrinos to mix with the muon and tau neutrinos. In
\fig~\ref{fig:tau-1au}, we show the yields at 1~AU for 250~GeV WIMPs
annihilating into $\tau^+ \tau^-$ both with and without neutrino
oscillations. Compared to the results at the surface of the Sun (\cf,
\fig~\ref{fig:tau-rsun}), we note that also the electron neutrinos are
effectively mixed with the muon and tau neutrinos. This effect is
solely due to vacuum neutrino oscillations governed by the small mass
squared difference $\Delta m_{21}^2$ (since $\theta_{13} = 0$). As can
be seen, the mixing is large, but not complete, which is due to the
leptonic mixing angle $\theta_{12}$ being large but not
maximal. Furthermore, the yields of muon and tau neutrinos now
essentially coincide due to the fast oscillations governed by the
neutrino oscillation parameters $\Delta m_{31}^2$ and $\theta_{23}$
(because of the $L/E$ dependence of the neutrino oscillation phase,
the phase is more sensitive to the neutrino energy $E$ at a distance
of 1~AU than at the surface of the Sun). When comparing the results
with and without oscillations, we can observe that the main impact of
the oscillations is to weaken, or even erase, asymmetries in the
initial yields through lepton mixing. As a result of this, any WIMP
candidate, which produces an excessive amount of tau neutrinos, will
have its resulting muon neutrino yield at a distance of 1~AU
increased (in this case by a factor of 3--4). Thus, it will be easier to perform indirect detection of
such a WIMP candidate than what may be naively expected if not
including neutrino oscillations. On the other hand, any WIMP
candidate, which produces an excessive amount of muon neutrinos, will
have its muon neutrino yield at a distance of 1~AU decreased and will
therefore be harder to detect. Furthermore, it is interesting to note
that, while electron and muon neutrinos are produced in equal amounts,
neutrino oscillations affect the yields in such a way as to have an
equal amount of muon and tau neutrinos in the later stages of
evolution.

\section{Propagation to and through the Earth to an actual detector}
\label{sec:propearth}

In the preceding section, we propagated the neutrinos to a distance of
1~AU from the Sun and we are now ready to proceed with the final
propagation to the Earth and through the Earth to an actual
detector. The reason to separate the propagation to 1~AU and the final
propagation to the detector is mainly practical. The propagation to
1~AU is neither time nor detector dependent, but the final propagation
to the actual position of the Earth and the detector is both time and
detector location dependent. Hence, we have split the propagation in
this way to make it easier to use the results of the previous section
for Monte Carlo simulations for actual detector responses.

In order to include these effects, we use a simple model of the
Earth's motion. This model contains a few approximations, but still
incorporates the interesting physics effects. We include the
eccentricity of the Earth's orbit and the daily rotation around its
own axis, but we make the simplifying assumption that both perihelion
and the winter solstice occurs at New Year (both are off by about one
week, but this approximation has no large effect on our
results). Within this model, we choose a time (fraction of the year
since New Year) of each event and for this time we calculate the
Earth's actual distance from the Sun. We then use this information to
propagate the neutrinos from 1~AU to the actual position of the Earth
(in case the Earth is closer than 1~AU, we just apply the inverse of
the propagation operator). In future releases of the {\sc WimpSim} code,
we may include a more sophisticated astronomical calculation. This has no effect on the results of this paper, since the errors introduced by these approximations are not observable. 

Furthermore, we have to determine the path length traversed through
the Earth. We perform this by specifying the latitude of
the detector and calculate the orientation of the Earth at the time of
the event. We use this knowledge to compute the path the neutrino will
take through the Earth. The propagation through the Earth is done in
the same way as through the Sun, \ie, we apply the neutrino evolution
operator method described in \Sec~\ref{sec:nuosc}, where we use the
Earth's matter density profile as given in \Ref~\cite{earthcomp} (from
which we calculate the electron density as a function of radius). We
then propagate the neutrino in steps of 0.3~\% of the radius of the
Earth, which turns out to be a good compromise between accuracy and
speed (the accuracy is far better than 1~\%). Even in this case, we make an astronomical calculation in order to simplify our simulation, which is that the detector is pointing towards the Sun and January 1 at 00:00. Again, this approximation has no effect on the results of our calculation.
\begin{figure}
\centerline{\epsfig{file=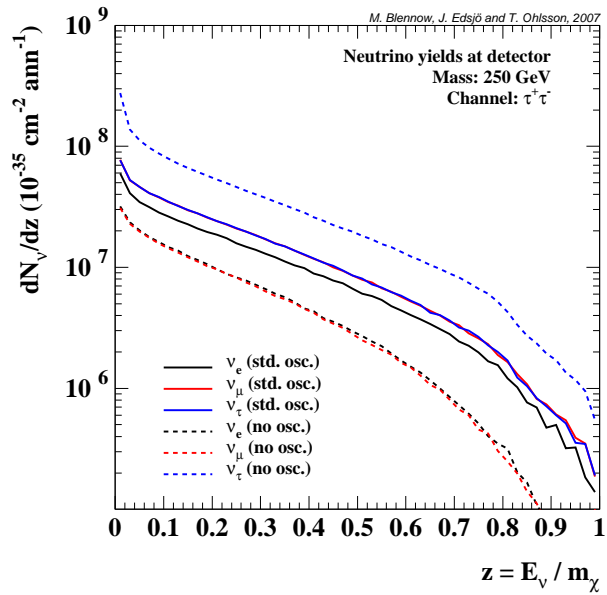,width=0.49\textwidth}}
\caption{The time-averaged neutrino yields as a function of $z =
E_\nu/m_\chi$ at the South Pole for annihilation in the Sun of 250~GeV
WIMPs into $\tau^+ \tau^-$ both with (std.~osc.) and without neutrino
oscillations.}
\label{fig:tau-det}
\end{figure}

Once we have propagated the neutrinos to the detector, we let them interact and again simulate their CC and/or NC interactions to produce hadronic showers and/or charged leptons. Both the neutrino yields at the detector as well as the hadronic showers and neutrino-induced lepton yields can be calculated. In case a charged muon is produced, we further let it traverse the detector medium where it can undergo both energy losses and multiple Coulomb-scattering to obtain the muon flux. In order to calculate the neutrino interactions and energy losses of muons properly, we need to specify what kind of medium the detector and its surroundings are made of. We have included both ice/water and rock as possible mediums. We do not show any results for the hadronic showers or charged leptons at the detector, since they contain no new information regarding the neutrino oscillation effects. Nevertheless, all these results are available at the website in \Ref~\cite{wimpsim}.

\subsection{Results for neutrino yields in an actual detector}

In order to be more specific, let us assume that the detector is at
$-90^\circ$ latitude (\eg~the IceCube detector \cite{icecube}
at the South Pole) 
and that the viewing period is from spring
equinox to autumn equinox, \ie, when the Sun is below the horizon for
a detector at the South Pole. For the same channel as before,
$\tau^+\tau^-$ from 250~GeV WIMPs, we show in \fig~\ref{fig:tau-det}
the neutrino yields at this detector, averaged over this viewing
period. The main effect is that the oscillations that were present in
previous figures are now almost completely washed out. It turns out
that the matter effects from the passage through the Earth are
negligible and the main reason for this ``averaging'' comes from the
eccentricity of the Earth's orbit, \ie, the variation of the Sun-Earth
distance effectively causes a wash-out of the oscillation pattern seen
at 1~AU. Of course, this is an artifact of our choice of averaging the
yield over the viewing period between the spring and autumn equinoxes
in \fig~\ref{fig:tau-det}. In addition, neutrino telescopes have a
rather poor energy resolution (50~\% or worse in this energy range),
and in practice, the oscillations in energy seen at a specific time
(and Earth-Sun distance) are not observable anyway.

However, for a source of monochromatic neutrinos (\eg~annihilation
directly to $\nu \bar{\nu}$), this effect could in principle be
observable, if the number of events is large enough. This would then
be seen as a variation of the event rate over time as the Earth would
move in and out of the oscillation phases. Nevertheless, for the
leading dark matter candidates, neutralinos in supersymmetric models
and Kaluza--Klein dark matter, this signature would not be
observable. The reason is that neutralinos do not annihilate to
$\nu\bar{\nu}$ directly, whereas for Kaluza--Klein dark matter that do
annihilate to $\nu \bar{\nu}$, they annihilate to all flavors with
equal branching ratios, and then, neutrino oscillations will not
affect the yields anyway.

\section{Non-zero $\boldsymbol{\theta_{13}}$ and comparison between normal and
inverted mass hierarchy}
\label{sec:nonzero_comparison}

In this section, we investigate the impact of a non-zero leptonic
mixing angle $\theta_{13}$ as well as the differences between normal
and inverted mass hierarchies on the neutrino yields from WIMP
annihilations. We also discuss the differences that appear when
considering antineutrinos. We present the neutrino yields at the
surface of the Sun as well as at the distance of 1~AU.

\begin{figure}
\begin{tabular}{cc}
\centerline{\epsfig{file=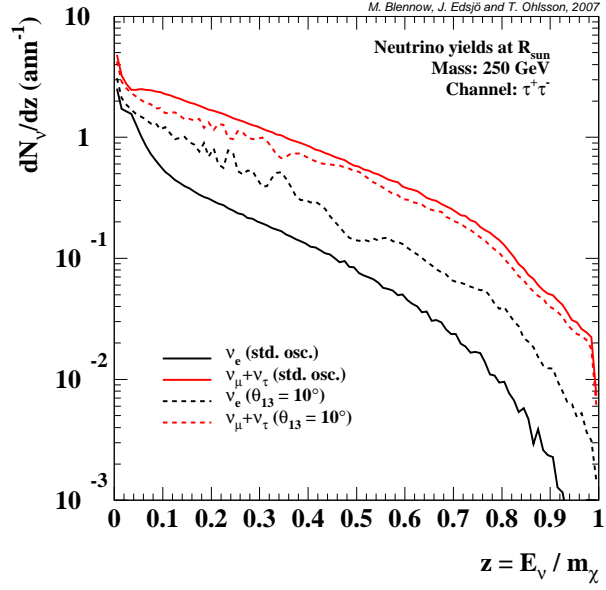,width=0.49\textwidth}}
\end{tabular}
\caption{The neutrino yields as a function of $z = E_\nu/m_\chi$ for
normal mass hierarchy at the surface of the Sun for annihilation of
250~GeV WIMPs into $\tau^+\tau^-$ for both $\theta_{13}= 0$
(std.~osc.) and $\theta_{13} = 10^\circ$.}
\label{fig:comp_rsun}
\end{figure}

In \fig~\ref{fig:comp_rsun}, we plot the neutrino yields for
$\theta_{13} = 0$ as well as for $\theta_{13} = 10^\circ$ and normal
mass hierarchy at the surface of the Sun.
We observe that, in the case of a non-zero $\theta_{13}$, the yield of
electron neutrinos is enhanced compared with the yields for muon and
tau neutrinos, which are accordingly suppressed. In fact, for
$\theta_{13} = 10^\circ$, the averages of the yields for different
neutrino flavors are comparable in magnitude, especially for low
neutrino energies (note that we plot the combined muon and tau
neutrino yields). The conversion of muon and tau neutrinos into
electron neutrinos for non-zero $\theta_{13}$ is due to adiabatic MSW
flavor conversions taking a larger role at the high MSW resonance. For
$\theta_{13} = 0$, adiabaticity (and thus also the flavor conversion)
is lost. In the case of inverted mass hierarchy, the yields for
$\theta_{13} = 0$ and $\theta_{13} = 10^\circ$ are qualitatively equal
to the yields for normal mass hierarchy and $\theta_{13} = 0$. This
means that a non-zero value of the mixing angle $\theta_{13}$ has no
effect for inverted mass hierarchy. In the case of antineutrinos, the
situation is the opposite, the yields for normal mass hierarchy are
qualitatively the same, whereas there is a difference between the
yields for inverted mass hierarchy. In this case, the reason is that
considering antineutrinos instead of neutrinos the replacement $V \to
- V$ has to be performed, and thus, there is a resonance for
antineutrinos only if $\Delta m_{31}^2 < 0$ (inverted mass
hierarchy). Naturally, the antineutrino yields for normal and inverted
mass hierarchies are qualitatively the same for $\theta_{13} = 0$.

\begin{figure}
\begin{tabular}{cc}
\centerline{\epsfig{file=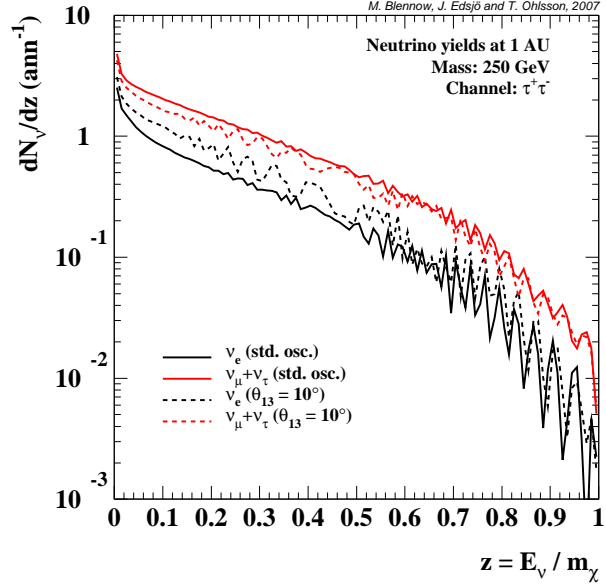,width=0.49\textwidth}}
\end{tabular}
\caption{The neutrino yields as a function of $z = E_\nu/m_\chi$ for
normal mass hierarchy at the distance of 1~AU for annihilation of
250~GeV WIMPs into $\tau^+\tau^-$ for both $\theta_{13}=0$ (std.~osc.)
and $\theta_{13} = 10^\circ$.}
\label{fig:comp_1AU}
\end{figure}

In \fig~\ref{fig:comp_1AU}, we plot the neutrino yields corresponding
to those in \fig~\ref{fig:comp_rsun}, but instead at the distance of
1~AU.
Again, we observe that in the case of normal mass hierarchy, the
neutrino yields for $\theta_{13} = 0$ and $\theta_{13} = 10^\circ$ are
different from each other. In the case of $\theta_{13} = 0$, the
electron neutrino yield is generally lower than the yields for muon
and tau neutrinos, whereas for $\theta_{13} = 10^\circ$, the electron
neutrino yield is generally higher than the yields for muon and tau
neutrinos. As in the case at the surface of the Sun and for low
neutrino energies, the averages of the yields of different neutrino
flavors are comparable in magnitude. However, in the case of inverted
mass hierarchy, the neutrino yields are essentially independent of the
mixing angle $\theta_{13}$, and the yields are similar to those in the
case of normal mass hierarchy and $\theta_{13} = 0$. In the case of
antineutrinos, the discussion of the yields at the distance 1~AU is
similar to the discussion for the yields at the surface of the Sun,
which means that effects on the yields appear for $\theta_{13} =
10^\circ$ and inverted mass hierarchy. In all other cases, the yields
are qualitatively equal.

We conclude that a non-zero leptonic mixing angle $\theta_{13}$ and
the neutrino mass hierarchy may influence the neutrino yields from
WIMP annihilations. Therefore, neutrino yields from WIMP annihilations
could, in principle, be used as a tool in determining both a non-zero
value of the mixing angle $\theta_{13}$ as well as the mass hierarchy
of the neutrino mass eigenstates. However, this would require the
detector to discriminate between neutrinos and antineutrino, which current neutrino telescope designs cannot do.

The effects of non-zero $\theta_{13}$ and the type of neutrino mass hierarchy have also been studied in \Ref~\cite{Lehnert:2007fv}. The neutrino flavor ratio results presented in \Ref~\cite{Lehnert:2007fv} are in agreement (up to the small effects due to the use of slightly different neutrino oscillation parameters) with our low-energy results, where neutrino interactions do not play a major role. For the high-energy part of the spectra, it is apparent that such flavor ratios cannot be computed accurately while neglecting neutrino interactions.

\section{Annihilation in the Earth}
\label{sec:anniEarth}

\begin{figure}
\begin{tabular}{cc}
\centerline{\epsfig{file=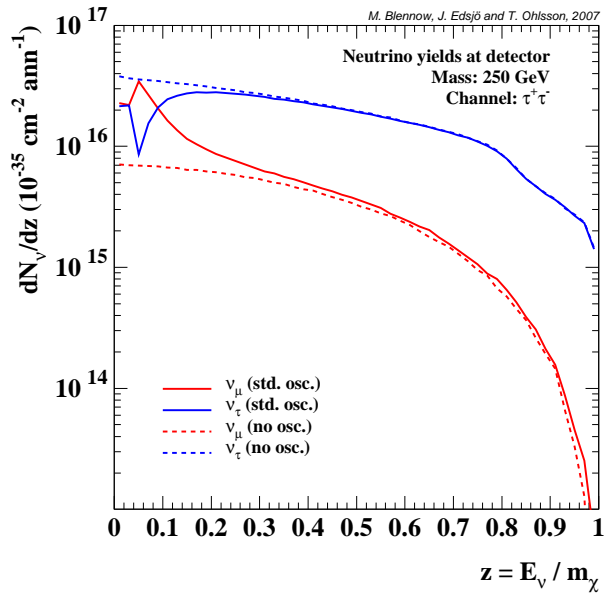,width=0.49\textwidth}}
\end{tabular}
\caption{The neutrino yields as a function of $z = E_\nu/m_\chi$ at
the surface of the Earth for annihilation of 250~GeV WIMPs into
$\tau^+\tau^-$ in the center of the Earth for the standard oscillation
parameters and without neutrino oscillations. The electron neutrino
yields are not plotted, since they (both with and without
oscillations) are identical to the muon neutrino yields without
neutrino oscillations.}
\label{fig:earth}
\end{figure}

In the case of WIMP annihilations in the Earth, the procedure is the
same as for the Sun, \ie, we let the neutrinos be produced around the
center of the Earth and then propagate them to the detector using the
same three-flavor framework as used for the Sun although there are a
few differences which we have to take into account. First, we must
consider the size of the annihilation region. This is fairly easy to
do given the temperature of the Earth's core and we follow the
procedure in \Ref~\cite{edgonprof} (which is really \eq~(\ref{eq:aprof1}) applied to the Earth). Second, since the density of the
Earth is much smaller than that of the Sun, $B$ meson interactions are
negligible. Third, neutrino interactions on the way out to
the surface are negligible for the energies of interest here. The geometry
is also simpler, since all neutrino telescopes are at the same
distance from the Earth's core. Hence, the only effect we need to take
into account after production around the Earth's core are neutrino
oscillations on the way to its surface. For each event, we simulate
the production point and let the neutrino propagate the actual path
length to the detector (using the Earth matter density profile from
\Ref~\cite{earthcomp}).

In \fig~\ref{fig:earth}, we show the yields of neutrinos at the
surface of the Earth for 250~GeV WIMPs annihilating into $\tau^+
\tau^-$ in the Earth's core. The effects of oscillations are seen as a
decrease of the tau neutrino yield and an increase of the muon
neutrino yield at low energies (below about 50~GeV). Thus, for higher
energies, that are of most interest to the currently planned neutrino
telescopes, neutrino oscillations do not have a large effect on the
neutrino yields from annihilations of WIMPs inside the Earth.

\section{Summary and conclusions}
\label{sec:conclusions}

In this paper, we have studied neutrinos originating from
annihilations of WIMPs inside the Sun and the Earth. We have computed
the yields of all neutrino flavors in a number of different WIMP
annihilation channels using an event-based full three-flavor Monte
Carlo, which we have made publicly available \cite{wimpsim}. Our Monte
Carlo includes both the effects of neutrino interactions and neutrino
oscillations. While our study is event-based, earlier studies
including neutrino oscillations
\cite{torino-osc,Lehnert:2007fv,Barger:2007xf} have been 
performed using a density matrix formalism for the neutrino
distributions. These two techniques are equivalent, but we have chosen the event-based technique as it is easier to implement as a WIMP Monte Carlo for neutrino telescopes.

We have studied the yields from different annihilation channels for
several WIMP masses. However, it would be necessary to study the
branching ratios and annihilation rates for a specific WIMP model in
order to make predictions on whether or not that particular model is
detectable in a neutrino telescope. In such a study, one would also
have to consider backgrounds to the WIMP neutrino signal, such as
high-energy neutrinos being produced in reactions when cosmic rays hit
the solar corona \cite{Hettlage:1999zr,Learned:2000sw,Fogli:2006jk}. The background
is of the order of a few neutrino events per year in a neutrino
telescope such as IceCube. These kind of studies are left for a future work.

Although our example channel in this paper has generally been the
annihilations of WIMPs into pairs of $\tau^+$ and $\tau^-$ for a WIMP
mass of 250~GeV, our simulations have been performed for 13 different
annihilation channels and 19 different WIMP masses in the range from
10~GeV to $10^5$~GeV. In addition, we have performed all of the
simulations for different values of the leptonic mixing angle
$\theta_{13}$, normal and inverted neutrino mass hierarchy, as well as
with neutrino oscillations turned off. Furthermore, our examples have
focused on the results for neutrinos and not antineutrinos, which are also
accounted for in our Monte Carlo. The full results of our simulations
are available at the website in \Ref~\cite{wimpsim}, both as extensive lists of plots of the results,
but also as computer-readable data files. In a coming release of DarkSUSY \cite{darksusy},
these data files will also be available in a simple-to-use way.

We have found that, for WIMP annihilations inside the Sun, the main
effects of neutrino oscillations are the effective mixing of muon and
tau neutrinos during the propagation to the solar surface and the
consequent mixing of electron neutrinos during the propagation from
the solar surface to a distance of 1~AU. However, due to the leptonic
mixing angle $\theta_{12}$ being less than maximal, the mixing between
electron neutrinos and the other neutrino flavors is not complete. As
a result, neutrino oscillations imply that the final yields of muon
and tau neutrinos are more or less equal at the Earth. This is in
contrast to the neutrino yields at production in the Sun, where the
electron and muon neutrino yields are typically equal, which is then also the
case at a distance of 1~AU if neutrino oscillations are ignored.

For non-zero $\theta_{13}$, we have found that the adiabatic flavor
transitions at the high MSW resonance results in the electron
neutrinos mixing with the muon and tau neutrinos already during the
evolution to the solar surface in the case of normal neutrino mass
hierarchy. For antineutrinos, we have the opposite situation, since
the high MSW resonance occurs for antineutrinos only when the neutrino
mass hierarchy is inverted. In the case of WIMP annihilations inside
the Earth, we have found that neutrino oscillations do not play any
major role except for at very low energies (less than about 50~GeV).

Our findings imply that, depending on the annihilation branching
ratios of the WIMP model under consideration, the yield of muon
neutrinos, which is the most interesting one from the neutrino
telescope point of view, can be either increased or decreased. For
example, in the case of neutralino dark matter, there is usually a
smaller amount of tau neutrinos produced. The effective flavor mixing
due to oscillations then implies that the muon neutrino yield will
decrease due to a larger number of muon neutrinos oscillating into tau
neutrinos than vice versa. On the other hand, dark matter consisting
of Kaluza--Klein excitations of Standard Model particles can have
large branching ratios into charged leptons. Out of these charged
leptons, only the taus decay before loosing a significant amount of
energy. Thus, there will be a larger production of tau neutrinos which
can subsequently oscillate into electron or muon neutrinos to actually
give a larger muon neutrino signal at a neutrino telescope.

\acknowledgments

We would like to thank M.~Cirelli and N.~Fornengo for the useful 
discussions we have had comparing our results.
This work was supported by the Royal Swedish Academy of Sciences (KVA)
[T.O.] and Swedish Research Council (Vetenskapsr{\aa}det), Contract
Nos.~622-2003-6025 [J.E.] and 621-2005-3588 [T.O.].

\appendix

\section{Equivalence of Monte Carlo and density matrix formalism}
\label{app:MCrho-equiv}

In order for our Monte Carlo simulations to make sense, it is
necessary to show that it is equivalent to the density matrix
formalism for neutrinos from WIMP annihilations presented in
\Ref~\cite{torino-osc}. At any point of propagation $r$, we can
construct a density matrix (normalized per annihilation) according to
\begin{equation}
\rho(r,E) = \frac{1}{N} \sum_a \ket{\nu_a}\bra{\nu_a} \delta(E-E_a),
\end{equation}
where $N$ is the number of annihilations simulated and $\ket{\nu_a}$
are all of the neutrino states at distance $r$ resulting from the
Monte Carlo simulations of these annihilations (note that the number
of neutrinos does not need to be $N$). A similar definition can be
made for the density matrix $\bar\rho(r,E)$ for antineutrinos. Since
the evolution equation is linear in the density matrix, it is
sufficient to consider the evolution of this density matrix for a pure
neutrino state. In this case, the density matrix takes the form (up to
normalization)
\begin{equation}
\rho(r,E) = \delta(E-E_0) \rho_0,
\end{equation}
where $E_0$ is the energy of the neutrino and $\rho_0$ is a matrix in
flavor space describing the flavor composition. In our Monte Carlo
simulations, the same state is described by the state vector $\ket\nu$
such that $\rho_0 = \ket\nu \bra\nu$. If disregarding the appearance
of lower-energy neutrino states due to interactions (although keeping
the degradation of the amplitude at $E_0$ due to interactions), then
the statistical evolution of the state $\ket\nu$ in our Monte Carlo
simulations is given by
\begin{equation}
\i\dd{\ket\nu}{r} = \left(H - \i\frac{\Gamma}{2}\right) \ket\nu,
\end{equation}
where $H$ is the neutrino flavor evolution Hamiltonian and $\Gamma =
\diag(\Gamma_e,\Gamma_\mu,\Gamma_\tau)$ is the interaction rate
matrix. From this follows that the statistical evolution of the
density matrix is
\begin{equation}
\dd{\rho}{r} = \left([H,\rho_0]-\frac \i 2 \{\Gamma,
\rho_0\}\right)\delta(E-E_0).
\end{equation}
However, the right-hand side of this equation still lacks the terms
from states which have lost energy due to NC interactions and from
states created due to secondary neutrinos from tau neutrino CC
interactions. The effect of both of these interactions is to add new
states into the collection of states from which we build our density
matrix. If the NC interaction rate for a neutrino of energy $E_0$ to
produce a neutrino of energy between $E$ and $E+\d E$ is
$\gamma(E_0,E)\d E$, then the probability of adding a neutrino state
in this energy range during the propagation of the small distance
$\Delta r$ is $\Delta r \gamma(E_0,E)\d E$. In addition, the added
state will have the same flavor composition as the original state,
\ie, $\rho_0$. Thus, it follows that
\begin{equation}
\left.\dd{\rho}{r}\right|_{\rm NC} = \gamma(E_0,E) \rho_0
\end{equation}
for the case of our single neutrino state.

For the regeneration of neutrinos with lower energies in CC
interactions, we note that neutrinos are only regenerated if the
interacting neutrino is a tau neutrino. If the interaction rate of tau
neutrinos at a neutrino energy of $E_0$ is $\Gamma_{\rm
CC}^\tau(E_0)$, then the rate of tau neutrino interaction for our
neutrino is $\rho_{0,\tau\tau}\Gamma_{\rm CC}^\tau(E_0)$, since
$\rho_{0,\tau\tau}$ is the probability for our neutrino to currently
be in the $\ket{\nu_\tau}$ state. If the energy distribution of the
new tau neutrino states is $f_\tau(E_0,E)$ and the energy distribution
of the new $\ket{\bar \nu_{e,\mu}}$ states is $f_{e,\mu}(E_0,E)$, then
we obtain
\begin{eqnarray}
\left.\dd{\rho}{r}\right|_{\rm CC} 
&=&
\Pi_{\tau} \rho_{0,\tau\tau} \Gamma_{\rm CC}^\tau(E_0) f_\tau(E_0,E),
\\
\left.\dd{\bar\rho}{r}\right|_{\rm CC} 
&=&
\bar\Pi_{e,\mu} \rho_{0,\tau\tau} \Gamma_{\rm CC}^\tau(E_0) f_{e,\mu}(E_0,E),
\end{eqnarray}
where $\Pi_\tau$ is a projector onto the $\ket{\nu_\tau}$ state and
$\bar\Pi_{e,\mu}$ is a projector onto the
$\ket{\bar\nu_e}$-$\ket{\bar\nu_\mu}$ subspace.

In total, we now have
\begin{eqnarray}
\dd{\rho}{r}(r,E) 
&=&
[H,\rho] - \frac \i 2 \{\Gamma,\rho\} + \gamma(E_0,E) \rho_0 +
\Pi_{\tau} \rho_{0,\tau\tau} \Gamma_{\rm CC}^\tau(E_0) f_\tau(E_0,E),
\\
\dd{\bar\rho}{r}(r,E) &=& \bar\Pi_{e,\mu} \rho_{0,\tau\tau}
\Gamma_{\rm CC}^\tau(E_0) f_{e,\mu}(E_0,E)
\end{eqnarray}
for $\rho(r,E) = \rho_0 \delta(E-E_0)$ and $\bar\rho(r,E) =
0$. Inserting this into the density matrix evolution equations of
\Ref~\cite{torino-osc}, we also obtain the very same evolution
equations, which proves that the two treatments are equivalent on a
statistical level.

\end{document}